\documentclass[twocolumn]{revtex4}
\usepackage{graphicx,epsfig}
\usepackage{amsmath}
\usepackage{amsfonts}
\usepackage[latin1]{inputenc}
\usepackage{natbib}

\usepackage{url}

\usepackage{fancyhdr}

\usepackage{mathtools}

\begin{document}

\pagestyle{fancy}

\title{\Large Addressing hard classical problems with Adiabatically Assisted Variational Quantum Eigensolvers}
\author{A. Garcia-Saez}
\affiliation{Barcelona Supercomputing Center (BSC), Barcelona, Spain.}
\author{J. I. Latorre}
\affiliation{Dept. F\'\i sica Qu\`antica i Astrof\'\i sica,  Universitat de Barcelona, Spain.}



\begin{abstract}
We present a hybrid classical-quantum algorithm to solve optimization problems in current quantum computers, 
whose basic idea is to assist variational quantum eigensolvers (VQE) with adiabatic change of the Hamiltonian. 
The rational for this new algorithm is to circumvent the problem of facing very small gradients in the classical optimization piece of a VQE,
while being able to run in current hardware efficient devices. 
A discrete concatenation of VQEs adapted to interpolating Hamiltonians provides a method to keep the quantum 
state always close to a path faithfully directed to find the final solution. We benchmark this Adiabatically Assisted Variational Quantum 
Eigensolver (AAVQE) on quantum Hamiltonians and hard classical problems, for which our approach shows fast convergence.
\end{abstract}


\maketitle


\section{Introduction}

The success of quantum computation relies on experimental improvements of quantum devices as well as on conceptual 
progress in the way quantum algorithms are designed. The increasing size of quantum computing prototypes provides enormous challenges to both 
research fronts. Within the superconducting qubit platform, new ideas to get better 
gate fidelity, less cross-talk among qubits and the possibility of adding error correction
strategies seem to pave a clear avenue for research \cite{scqbits}. It is necessary to match this 
experimental progress with new ideas on the way quantum computers can be used to solve
all sort of problems, both in physics and in real world applications \cite{reviewQC}.

Quantum algorithms were first developed to produce explicit quantum circuits that solved a task, 
often related to classical computation. Two of the most relevant instances of this original 
approach were the algorithms due to Grover \cite{grover} and Shor \cite{shor}. Both of them 
exploit the advantage of quantum superposition and parallel processing using quantum circuits. 
It is, thus, necessary to design specific circuits that take an initial banal quantum state 
into a final state which encodes the  solution to a classical problem. There are not many
quantum circuits that serve the purpose to explicitly solve quantum problems. Some exceptions 
are the ones associated to quantum systems which are integrable \cite{vcl,k}, and have been 
verified on actual quantum computers \cite{halk}. It is also possible to design a quantum 
circuit for the Kitaev honeycomb model \cite{so}.

A separate class of interest consists of quantum algorithms that solve a task, such
as optimization. The original idea of adiabatic quantum computation \cite{fggs,al} is a good example. 
The ground state of a given Hamiltonian can be obtained by first starting from the ground 
state of a simple Hamiltonian under control. Then the adiabatic theorem states that the two
ground states can be connected by evolving with a Hamiltonian that keeps changing in time. 
To be precise, let us consider a system with $n$ qubits that evolves with the Hamiltonian
\begin{equation}
  \label{adiabatic}
  H(s)=(1-s) H_0 + s H_P\quad ,
\end{equation}
where the first Hamiltonian can be chosen to have the simple product structure
\begin{equation}
  \label{H0}
  H_0=\sum_{i=1,\ldots,n} \sigma_i^x ,
\end{equation}
and $H_P$ is a problem Hamiltonian whose ground state encodes the problem solution. The parameter $s$ 
can be made to change in time from 0 to 1, that is, $s=s(t/T)$, being $T$ the total running time of 
the evolution. Given this conditions, the adiabatic theorem states that the initial ground state of 
$H_0$ will evolve to the ground state of $H_P$, providing the solution to the problem, if the evolution 
remains slow enough. To be precise, the probability error is given by
\begin{equation}
  \label{adiabaticcondition}
  P(error)=\max_s \frac{|\langle \psi_1 | \frac{dH(s)}{ds} | \psi_0\rangle|}{g(s)^2},
\end{equation}
where the numerator corresponds to the transition amplitude between the ground state 
and the first excited state, and the denominator is the gap of the system. The theorem
indicates that the source of error is related to the possibility of jumping from the 
ground state to the first excited state, either because the amplitude for such a process
is large or due a very small gap. In this language, hard problems are associated
to adiabatic evolutions where the gap becomes exponentially small.

A new category of algorithms has been put forward under the name of Variational Quantum Eigensolvers (VQE) 
\cite{vqe1,vqe2, vqe3, vqe4}. The basic idea is to consider quantum circuits which can be parametrically constructed
 from a reduced set of gates. By proper tunning of the circuit parameters we construct 
a variational ansatz to a problem under consideration. The characterization of such
a quantum circuit, that is the prescription of gates and the order in which they act, 
is purely classical. Then a hybrid algorithm can be constructed so that the classical
characterization of the quantum circuit is subject to a numerical optimization. The circuit
keeps improving iteratively until the variational ansatz provides a good
solution. This class of algorithms, based on hardware efficient circuits, 
has been used notably for quantum chemistry \cite{vqeIBM}.

In this wowrk we present an extension of VQE algorithms based on adiabatic evolution. 
We coin the name Adiabatically Assisted Variational Quantum Eigensolvers (AAVQE) for this classical strategy. 
We first present the idea in the next section, and then proceed to benchmark it. Some conclusions close the paper.


\section{The AAVQE algorithm}

Quantum algorithms should bring a quantum system from a state which is easy to prepare to another state which 
encodes the solution to a problem. An obvious difficulty for any quantum algorithm is how to find a fast and 
reliable way to connect both states, given the exponential size of the Hilbert space. As mentioned previously, 
a relevant class of multi-purpose quantum algorithms is the one labeled as VQE. 
The central idea is to use a parametrized quantum circuit as a provider of variational ansatz to find the ground state of a Hamiltonian. 
The algorithm proceeds by using a classical characterization of the quantum circuit which is then explored using optimization techniques. 
The class of VQE is an example of hybrid algorithms that
try improve the potential of quantum circuits with classical assistance.

This idea faces the problem of finding a reasonable path in the parameter space of the circuit to end up with the right solution. 
Notice however that it is possible to strongly argue against this strategy in the following way. Given the exponential size of the Hilbert space, 
any technique that searches for paths in the parameter space that characterizes the quantum circuit is bound to deal with very tiny gradients. 
These gradients can even be exponentially small. In such a case, the algorithm may not find the right gradient
and would be shooting around in a random way. No convergence to a good result would be
seen, specially for large problems \cite{googleML}. 

It is clear that adiabatic evolution gives a guaranteed path to find the ground state of a Hamiltonian, 
but this procedure may need a very slow evolution. It is also true that VQE
may get lost in the search of a minimum. We may summarize the pros and cons of both
methods  in the following way

\begin{itemize}
\item Variational Quantum Eigensolver
  \begin{itemize}
  \item PRO: Uses an arbitrary quantum circuit that is described by a set of parameters to generate variational a ansatz to minimize the problem Hamiltonian.
  \item PRO: Searches for the gradient in parameter space using classical optimization.
  \item CON: May get lost in parameter space.
  \item CON: Uses a large number of measurements.
  \end{itemize}
\item Adiabatic Quantum Evolution
  \begin{itemize}
  \item PRO: Always finds its way to the solution.
  \item PRO: Can be optimized to pick a more efficient adiabatic path.
  \item CON: May be exponentially slow.
\end{itemize}
\end{itemize}

\begin{figure}[h!]
\centering
\includegraphics[scale=0.45]{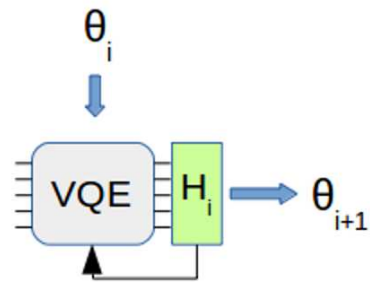}
\caption{Scheme for the AAVQE algorithm. The Hamiltonian problem is encoded as the end point of 
a discrete  adiabatic interpolation, $H(s)=(1-s) H_0+s H_P$, where $s$ takes values between 0 and 1, 
that is $s=0,s_1,s_2,\ldots,s_T$ and $s_T=1$. At each stage of
the computation a VQE solves for the ground state of the intermediate Hamiltonians. The arrows 
represent these classical optimization processes. After minimization at step $s_i$, the characterization 
of the optimized quantum circuit is used as the initial parameters for the $s_{i+1}$ minimization. 
The Hamiltonian is then changed and a new VQE is operated.
}\label{fig:aavqe}
\end{figure}

The new idea we here put forward is to combine the virtues of adiabatic evolution with those of the VQE strategy. 
We coin the term Adiabatically Assisted Variational Quantum Eigensolvers (AAVQE) for these class of algorithms. 
The basic idea is represented in Fig. \ref{fig:aavqe}. The algorithm works by applying VQE to a 
series of Hamiltonians that keep evolving from a simple one to the problem we need to solve.

Let us make the algorithm concrete. We first create a Hamiltonian suited for adiabatic evolution 
following the form of Eq. \ref{adiabatic}. This Hamiltonian, though, will
never be used for any real time evolution. Instead, we shall consider each discrete step
as a new problem to be solved using a VQE. At every step of the computation, the VQE needs the
quantum circuit to be initialized with a classical set of parameters $\theta^{(0)}_{s_i}$ (initially
this set can be chosen randomly). After
minimization is complete the quantum circuit is defined by the trained final set of parameters 
$\theta^{(f)}_{s_i}$. We then to use as initial parameters those obtained in the 
previous step. This means that the final parameters that deliver the optimal minimum for the Hamiltonian at any step $s_i$ are
passed as initial parameters for the next one, that is $\theta^{(f)}_{s_i}=\theta^{(0)}_{s_{i+1}}$.

There are two main differences of our algorithm  with respect to the traditional adiabatic
evolution, each one serving adiabatic evolution and VQE in both directions.
The first one is that no real time evolution is made to get to the next correct ground state. 
A discretized adiabatic evolution is just used to guide a series of VQEs to find
its way to the final Hamiltonian groundstate.
The second is that VQE provides a better groundstate to be used at the subsequent step in the 
adiabatic evolution series. The role of the adiabatic interpolation is simply to guide the series of VQE that deliver a flow of
 ground states  towards the final solution. The summary of the algorithm reads:

\begin{enumerate}
\item Prepare the ground state of a simple Hamiltonian with $s_0=0$ with a quantum circuit using VQE. Its final characterization is given by the parameters $\theta^{(f)}_{s_0}$
\item Add a step $\Delta s$, that is $s_{i+1}=s_i+\Delta s$.
\item Run a VQE on $H(s_{i+1})$ using as initial parameters $\theta^{(0)}_{s_{i+1}}=\theta^{(f)}_{s_i}$, the final  parameters from the previous step.
\item If $s=1$ stop, else go to 2).
\end{enumerate}

The combination of adiabatic optimization and variational eigensolvers has been explored before, i.e. Ref.\cite{vqe2} where the particular choice of the adiabatic is the subject of a VQE optimization. The final state preparation improves as a particular path among a parametrization is selected. 
In the examples presented in our work, this path is fixed. In Ref.\cite{tpvqe},
an adiabatic strategy is used over the QAOA algorithm \cite{qaoa}. The Trotter parameters are the subject of the optimization, and are evolved following an adiabatic transformation similar to the one presented here. Our work makes no use of a particular Hamiltonian evolution and the quantum circuit employes has no relation to the problem hamiltonian. This leads to the application of the AAVQE to problems without an
explicit implementation of the circuit Hamiltonian, such as classical
optimization problems, making quantum optimization available to any quantum circuit defined by classical parameters. 
In the Results section below we show that precisely these problems benefit specially from an adiabatic approach.

There are two main tunable options in the AAVQE algorithm. First, the detailed discretization
of the adiabatic change of $H(s)$ is a matter of choice. The more steps we use, the easier will be for the algorithm to remain in the ground state but more calls to the quantum circuit will be needed. The scaling of the running cost of the algorithm is 
linearly proportional to this discretization. If $s$ is divided in $T$ steps,
the number of quantum gates which are needed grows as $T$. It is also possible to look
for optimal discretization of adiabatic evolution which are not linear.
Second, the
classical optimization method used to find the gradient towards the ground state
is also a choice in the hands of the programmer. If we consider simple methods
based on computations of local gradients, it is then a matter of how this gradient is computed. 
All in all the algorithm has a good amount of freedom we shall explore
in the next chapter.

Let us note that AAVQE works for any kind of final problem, let it be classical or quantum. We shall explore both cases later on. 
In practice, each VQE needs to minimize a function which is
obtained as measurement on the state produced by the circuit. In the case that the minimization function can be read 
directly from the output probabilities in each qubit in the computational basis, the problem is essentially equivalent to a diagonal 
Hamiltonian. If, instead, we optimize a Hamiltonian involving many-body operators terms, expectation values of composite operators will form 
the fit function used by the classical optimization part of the algorithm. 

There is a further relevant conceptual advantage for AAVQE over VQE, which is related to the 
possibility of VQE to get trapped in a local minimum. That is, VQE, being a plain 
minimization strategy, can converge to a quantum circuit which is not delivering the 
state with an absolute minimum of energy. Instead, adiabatic evolution may be slow but
it will go to the right minimum. This is to be taken with a grain of salt, as adiabatic
evolution may end up jumping to an excited state if it runs too fast. Still, adiabatically assisted 
eigensolvers may have the right balance to avoid local minima, as we shall see in our benchmarking.


\section{Results}

In our experiments we simulate a quantum circuit controlled by a classical optimization algorithm. We follow a practical 
approach based on current technology,
using design conditions similar to those available in experimental labs (see \cite{vqeIBM}). This is reflected both in 
the quantum circuits we optimize, and in the classical numerical methods currently used.

We simulate the quantum circuit and the classical optimization without any noise effect. Having access to the full wavefunction desciption we may use a full estimation of the energy, or rather simulate the 
error introduced by quantum measurements. We apply these techniques to a purely quantum problem, namely the XX spin chain with local field 
$H=\sum \sigma_x \sigma_x + \lambda \sigma_x$, and also  
to hard instances of a hard classical combinatorial 
problem, namely the EXACT COVER problem. As with VQE, the quantum circuit is exactly the same for 
all problems, the only difference in implementation being the classical evaluation of the objective function.

\subsection{Optimization of quantum problems}

Experimental implementations of the VQE method have been successfully applied to Chemical problems
\cite{vqeIBM}. The classical optimization in this setup is performed by the SPSA algorithm \cite{SPSA,SPSAtomo}, 
which offers an efficient procedure to obtain the ground state of a quantum model 
using a reduced number of energy evaluations. The method provides an estimate of the gradient
function around the evaluation point, and this gradient is used to decide the optimization direction
in the following evaluation of the energy.

\begin{figure}[ht]
\centering
\includegraphics[scale=0.3]{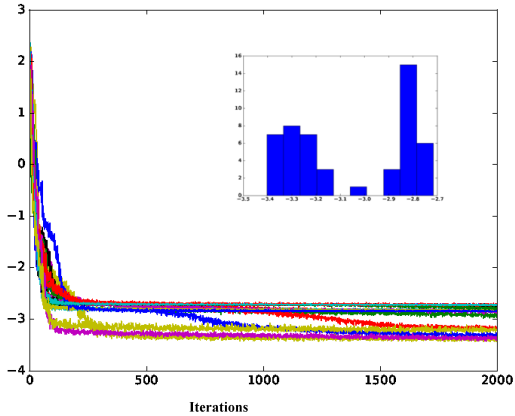}
\caption{We simulate the run of a set of $20$ optimizations of the ground state of the XX spin chain
using the VQE optimization. The initial state is chosen randomly, but it is equal in all simulations. While the method
is capable of reaching a value close to the ground state, a typical run may get stuck in local minima, even at 
very long optimization times. The inset shows the energy distribution of $100$ different runs after $1000$ 
optimization steps.}\label{fig:4spsa}
\end{figure}

While the SPSA efficiently solves local problems, the optimization relies solely on geometric properties of the evaluation function. 
Thus, for some instances it may face problems to determine a valid optimization direction, resulting in local minima. As the evolution of 
the optimization of the SPSA method tries to fine grain a good solution, escaping the local minima gets harder at final stages of the optimization. 
As the method is based on a random choice of the gradient evaluation points, different runs of the optimization may result in different final values of the energy.

\begin{figure}[ht]
\centering
\includegraphics[scale=0.21]{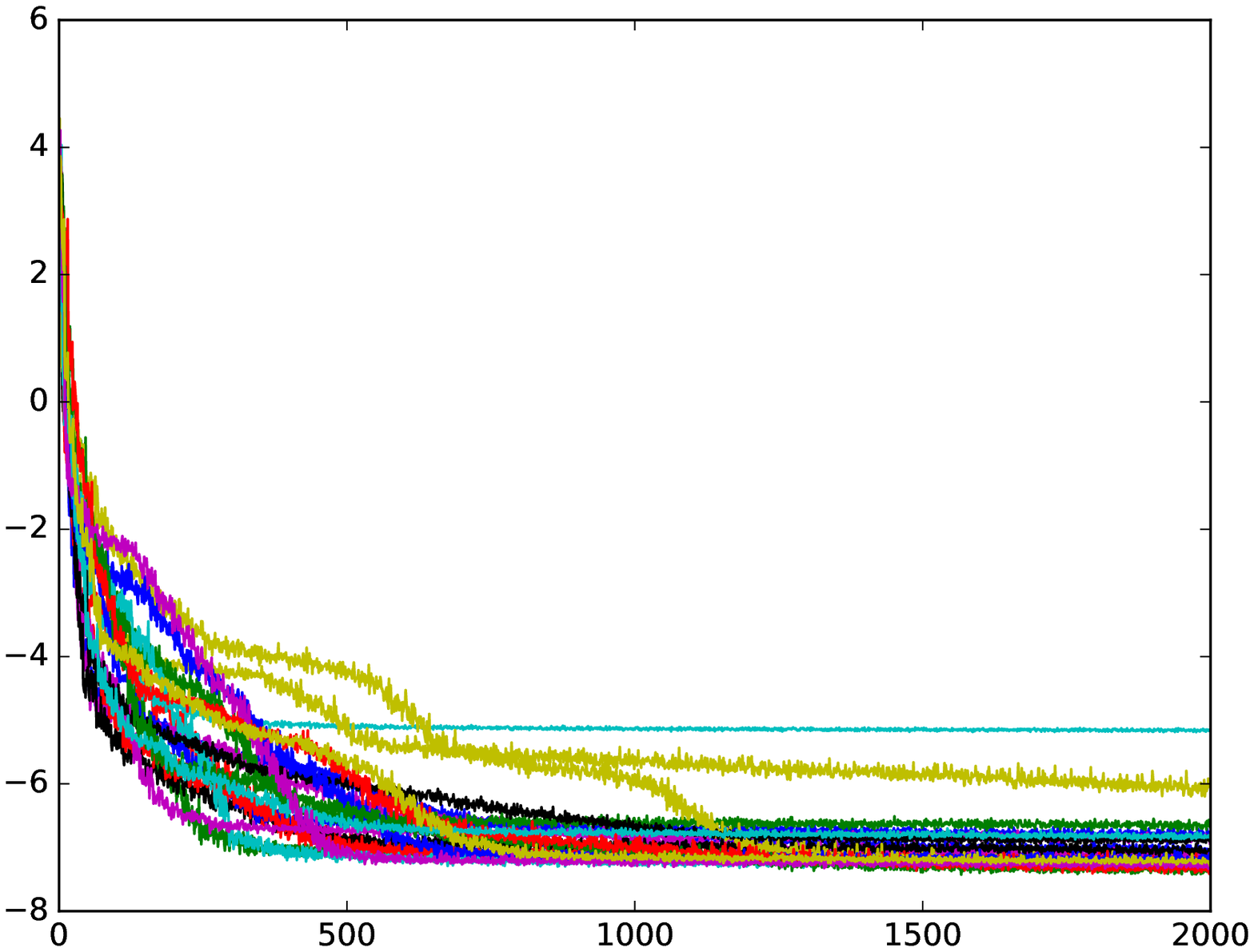}
\includegraphics[scale=0.21]{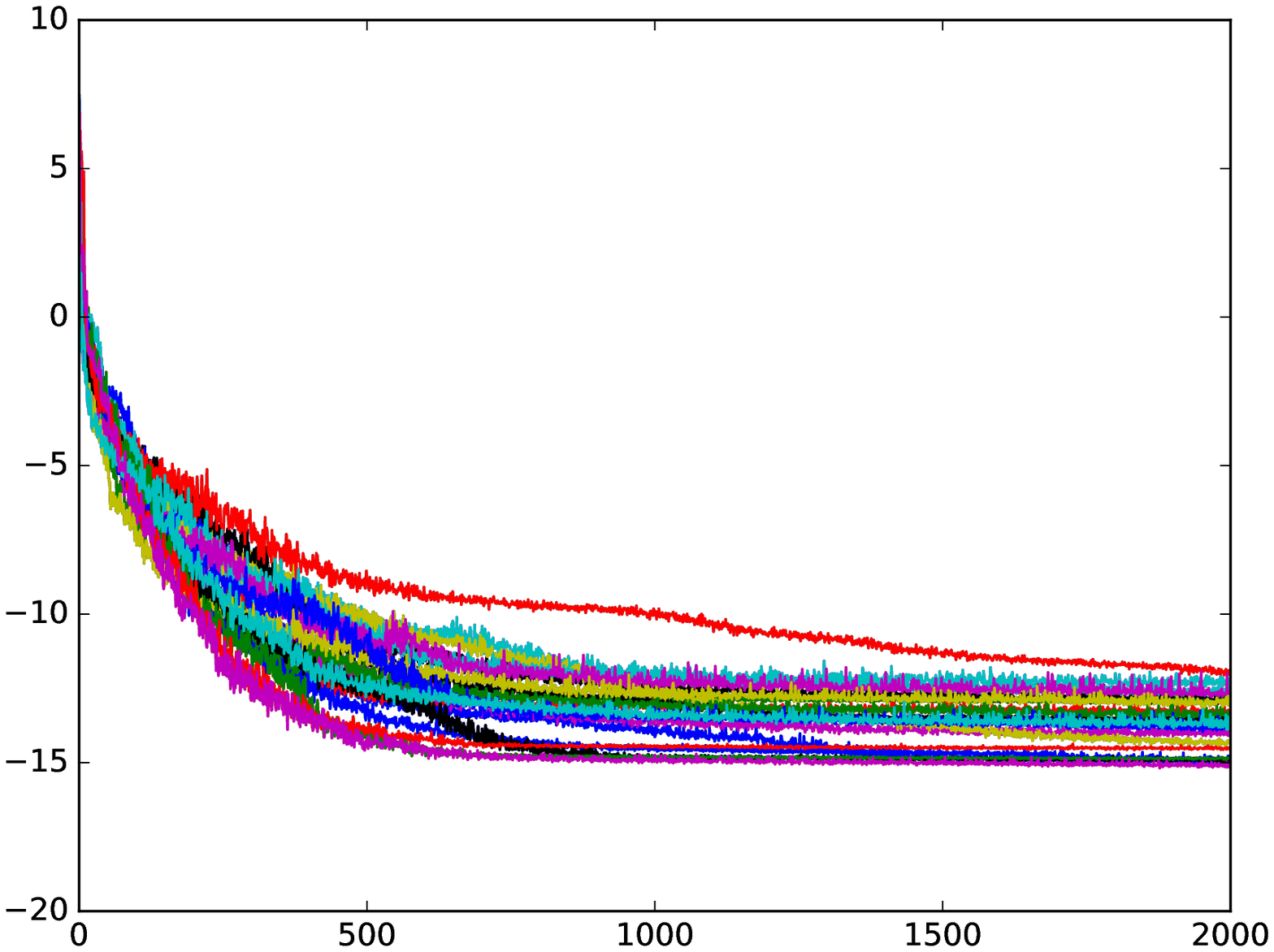}
\caption{Same as depicted in Fig. \ref{fig:4spsa} with systems of size $N=8$ (left) and $N=16$ (right).
The presence of local minima persists, and the final error gets significantly larger.}\label{fig:8spsa}
\end{figure}

The behavior of the VQE implementation using SPSA is shown in Fig. \ref{fig:4spsa} for $N=4$. While at initial stages of the optimization the numerical evolution performs well and the energy approaches rapidly the ground state value, the SPSA may easily get stuck in local minima far 
from the ground state value. More importantly, this situation may persist even at very long optimization times. The situation appears 
in the general case of longer chains and different values of the local field (see Fig. \ref{fig:8spsa}).
The introduction of $H(s)$ in the general minimization problem allows a fine control of the optimization. 
After a discretization of the $s = \{0,1\}$ interval, a succession of optimization problems are set according to $H(s)$. 
An independent optimization of the ground state is performed for each value of $s$, using the relation $\theta^{(f)}_{s_i}=\theta^{(0)}_{s_{i+1}}$.

\begin{figure}[ht]
\centering
\includegraphics[scale=0.35]{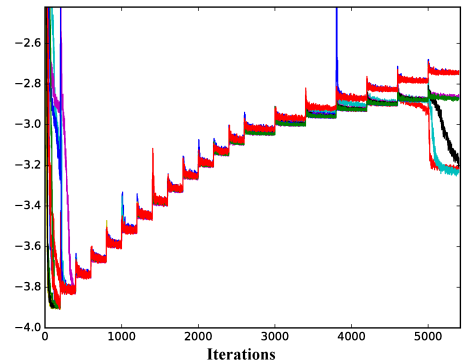}
\caption{The result of running the complete adiabatic transition using the AAVQE algorithm on a spin chain. We use $\Delta=0.05$ and SPSA for the VQE computation.
While we observe convergence to the correct value, some instances fail to converge in the last stages. In this region energy levels for this systems lie closer, so 
convergence of the method is harder. This can be easily solved running only the last stages using the parameters from the last optimization as an initial point.}\label{fig:4aavqe}
\end{figure}

The total evolution of the AAVQE algorithm for an XX spin chain with $N=4$ is shown in Fig.\ref{fig:4aavqe}. We plot the energy values 
--as evaluated by the quantum circuit-- at each intermediate evolution of the VQE solution, for each $H(s)$. Here we observe the effect of successive 
changes in $H(s)$, as the initial $\theta^{(0)}_s$ does not correspond to the ground state of the new Hamiltonian, producing an energy pike. 
This is rapidly corrected by the new optimization, which produces a new ground state for $H(s_{i+1})$. 
The final ground state encodes the solution to the target problem.

Our implementation of the AAVQE may also get stuck in local minima as we rely on the same classical methods as the VQE algorithm. However, there is a number of benefits of using the AAVQE besides those provided directly by the adiabatic theorem. For each $H(s_i)$ we may stop the optimization upon convergence, 
reducing the numerical effort for easy values of $s_i$. Upon failure to converge, we may restart the optimization at our 
best solution of a near problem $H(s_i)$, as we may efficiently store each previous solution. It is easy to detect hard 
regions of the adiabatic transition, as these normally yield different converged values of the energy (see i.e. Fig. \ref{fig:4aavqe}). 
On these regions, one may tune the optimization hyperparameters to carefully converge under harder conditions. We have fixed these parameters in our simulations,
so one may only expect better results with a fine parameter tunning.

\subsection{Optimization of classical problems}

Frustration-free quantum problems similar to the XX spin chain of the previous section show in general good convergence properties. 
After a few iterations of the VQE algorithm we may already have a good estimation of the ground state energy. 
The energy landscape seems to be smooth, and the AAVQE benefits showing good convergence far from regions 
with a closing energy gap. In this section, we test the AAVQE method on a family of hard classical problems. 
This scenario shows significant differences with the quantum Hamiltonians explored in the previous section, 
as frustration may play an important role on the solution space, 
resulting in harder optimization problems. 

We choose instances of the classical EXACT COVER problem --an NP-complete problem--, where a collection 
of sets of variables is evaluated. Each set, formed by 3 binary variables, 
is evaluated to true whenever a single variable of the set is $1$. A valid assignment satisfies this 
condition for each set of the collection.
We select hard instances for our simulations to stress the exploration of the AAVQE. 
These instances have a single valid assignment, and are hard as the 
particular choice of clauses requires a full exploration of the combinatorial possibilities.

This problem may be formulated in a Hamiltonian form using the diagonal expression
\begin{equation}
  \label{exactcover}
	H_{EC} = \sum_{<i,j,k>} (Z_i + Z_j + Z_k -1)^2
\end{equation}
with local operators $Z_i=\frac{1}{2}(\sigma_i-1)$. A correct assignment of classical values yields a minimal ground state eneregy $E_0 = 0$. 
Each set in the EXACT COVER instance translates into a term in the sum in Eq.\ref{exactcover}. A non-satisfying assignment will
violate a number of terms in the sum, contributing to the final value of the energy. The minimal gap is therefore $1$.
The optimization of the ground state may face situations where similar states (as measured i.e. by the Hamming distance)
may show very different energies, with effects in a local space exploration of the solutions.

\begin{figure}[ht]
\centering
\includegraphics[scale=0.4]{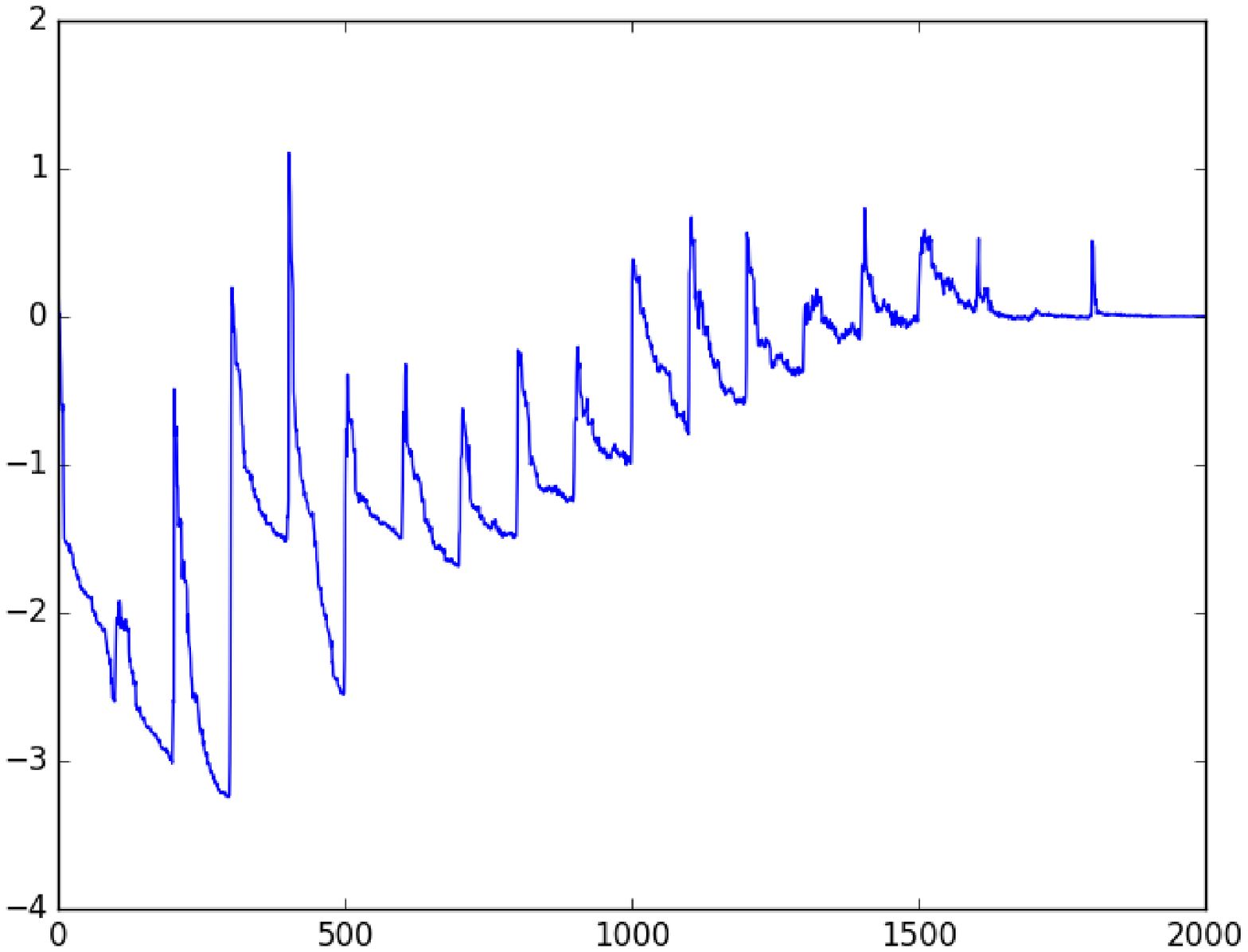}
\includegraphics[scale=0.4]{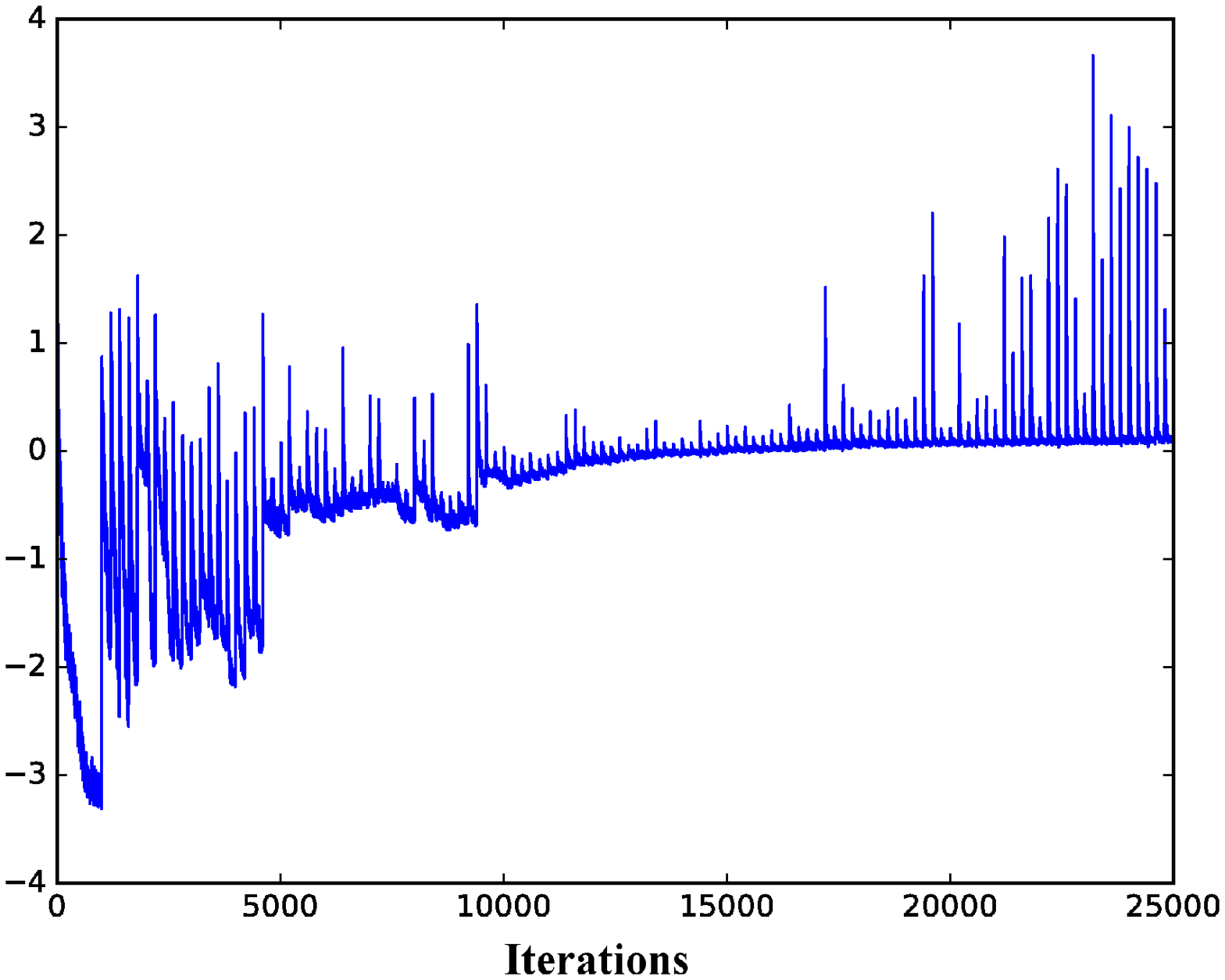}
\caption{The AAVQE algorithm is used to obtain solutions of hard instances of the EXACT COVER problem. We solve problems with $N=4$ (top)
and $N=8$ (bottom). While the final state encodes the correct solution, intermediate steps may deviate 
from the ground state of $H(s)$ as the energy spectrum may pre. Interestingly, successive steps of the VQE recover the ground state }\label{fig:maxcut4_8}
\end{figure}

We present in Fig. \ref{fig:maxcut4_8} results of the AAVQE on hard instances of the EXACT COVER problem for $N=4,8$.
We find the correct assignment of variables at the end of the adiabatic transition. To further stress the performance of AAVQE, we obtained the energy
after simulating the effect of a finite number of measurements of the quantum state. Even so, the AAVQE delivers a valid assignment of the variables.
While during the adiabatic evolution some partial optimizations may deviate of the ground state, the AAVQE is robust enough to return to the ground state in successive optimizations.

\begin{figure}[ht]
\centering
\includegraphics[scale=0.4]{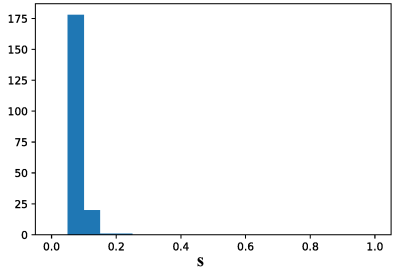}
\caption{We compute the adiabatic evolution of $200$ hard instances with $N=16$. This histogram shows the value of $s$ at which the correct solution is already present among the first rounds of intermediate measurements.
We use $100$ iterations of the VQE for each values of $s$ (with $\Delta s = 0.05$). Almost all instances are solved in less than $100$ VQE iterations.}\label{fig:s_hist_16}
\end{figure}

The adiabatic process evaluates the energy of the trial state at each optimization step. We have shown results of classical problems simulating real measurements. Each of these 
measurements contains detailed information of the evaluated function. As we look for a solution in the computational basis, one may find the solution even if the estimated ground state 
lies far from the ground state of $H_{EC}$. We plot the success of detecting the solution using only real measurements on 1000 iterations in Fig.\ref{fig:s_hist_16} for $N=16$, 
for which we normally find the correct solution in less than $100$ iterations and $s<0.1$. Similar results are obtained up to $N=20$. As overlap probabilities 
may be exponentially small for large problems, one may still find good approximate solutions as these are also preferred by the stochastic minimization. However, finding those does not
provide any information about the correct solution.

The quantum circuit and the parameter set used in these 
calculations is exactly identical to that used in the previous section for purely quantum problems, showing the flexibility the AAVQE shows for solving optimization problems. For the same instances, we were unable to find a valid set of parameters of the VQE to solve --or even approximate-- these problems (for a detailed exploration see Ref.\cite{hybridComb}). As further tunning of the parameters may solve this limitation, this additional numerical effort should be considered while evaluating the performance of the VQE method compared to the AAVQE algorithm.


\section{Conclusion}

The class of algorithms called Variational Quantum Eigensolvers can be refined adding
an adiabatic strategy to guide the initial state entering a discretized series of eigensolvers, what we call Adiabatically Assisted Variational Quantum Eigensolvers (AAVQE). 
We have benchmarked this new algorithm against VQE on the task of constructing the ground state of condensed matter systems. For quantum problems such as quantum spin chains,
where VQE performs well, the AAVQE shows similar results, but allowing extra parameter tunning.

We report also results of the AAVQE on hard instances of an NP-complete problem. In this scenario we are still able to obtain
good results where standard setting of the VQE algorithm fails.
The details of the implementation of gradient methods are critical for the performance of
both VQE and AAVQE. A more thorough benchmarking is still needed.

By design, the AAVQE allows for a more intelligent adiabatic interpolation. We have restricted our study to a simple transition
obtaining already good results. Further analysis should considerate additional optimization of transition trajectories and weight of clauses.

It is arguable that AAVQE may have an even advantage over VQE as the system size gets larger.
In that case, the problem of finding a good path to the final solution gets exponentially worse,
so an adiabatic strategy may be mandatory.


\section*{Acknowledgements}

We acknowledge funding from projects FIS2015-69167-C2-2-P and FIS2017-89860-P (MINECO/AEI/FEDER, UE).



\end{document}